\documentclass[12pt,a4paper]{article}

\usepackage{amsmath,amsfonts,graphics,epsfig,amssymb}
  \def\GeV{\,{\rm GeV}} 
\def\EeV{\,{\rm EeV}} \def\GeV2{\,{\rm GeV^2}} \def\GeV{\,{\rm GeV}} \def\mb{\,{\rm \mu{b}}}

\def\G{\,{\rm G}}  
     \def\Mpc{\,{\rm Mpc}}  
   \def\cmm2{{\,\rm 
cm^{-2}}} \def\cm2{{\,{\rm cm}^2}} \def\cmm3{{\,{\rm cm}^{-3}}} 
\def\gcmm3{{\,{\rm g\,cm^{-3}}}}

 \def\ga{\mathrel{\mathpalette\fun >}} 
\def\fun#1#2{\lower3.6pt\vbox{\baselineskip0pt\lineskip.9pt
  \ialign{$\mathsurround=0pt#1\hfil##\hfil$\crcr#2\crcr\sim\crcr}}}

\begin{document}
\title{ Double Pair Production by Ultra High Energy Cosmic Ray Photons } 
\author{ S.V.~Demidov\footnote{{\bf e-mail}: demidov@ms2.inr.ac.ru}, 
O.E.~Kalashev\footnote{{\bf e-mail}: kalashev@ms2.inr.ac.ru} \\
\small \em Institute for Nuclear Research of the Russian Academy of Sciences,\\
\small \em 60th October Anniversary Prospect 7a, Moscow 117312 Russia }
\date{}
\maketitle
\vspace{-12mm}
\begin{abstract}
With use of CompHEP package we've made the detailed estimate of the influence of 
double $e^+e^-$ pair production (DPP) by photons on the propagation of ultra 
high energy electromagnetic (EM) cascade. We show that in the models in which 
cosmic ray 
photons energy reaches few$\times10^3$~EeV refined DPP analysis may lead to substantial 
difference in predicted photon spectrum compared to previous rough estimates.
\end{abstract}

\section{Introduction}
Ultra-high energy (UHE) photons have not been recognized so far by any of present 
generation experiments~\cite{AGASA_eest,Yakutsk_spec,HiRes_spec,Auger_spec}, 
although their existence is predicted by Greisen--Zatsepin--Kuzmin 
effect~\cite{g,zk} as well as by most of hypothetical top-down models of UHE 
cosmic rays origin. There are several bounds on fraction and flux of 
ultra-high energy photons above $10 - 100\EeV$ obtained by independent 
experiments~\cite{A+Y,Ylim,Auger_sdlim}. Photon limits are used to constrain the 
parameters of top-down models (see for example~\cite{Gelmini:2005wu}). Future 
bounds may also limit considerable part of parameter space of astrophysical 
models, in which 
photons are produced as secondaries from interactions of primary protons or 
nuclei with cosmic microwave background (CMB). Understanding interactions of 
UHE photons with universal backgrounds is a crucial point for building such 
constraints.

In the wide energy range the spectra of electron and photon components of cosmic rays 
follow each other due to relatively rapid processes transferring $\gamma$-rays 
to electrons and backwards. Pair production (PP) and inverse Compton scattering 
(ICS) are the main processes that drive the EM cascade. In the Klein--Nishina 
limit where $s \gg m_e^2$, either electron or positron produced in a 
pair production event typically carries almost all of the initial total energy. 
The produced electron (positron) then undergoes ICS losing more than 90~\% of 
energy and finally the background photon carries away almost all of the 
initial energy of the UHE photon. Due to this cycle the energy loss rate of the 
leading particle in the EM cascade is more than one order of magnitude less 
than interaction rate. However in presence of a random extragalactic magnetic 
field (EGMF) the electrons may lose substantial part of their energy by 
emitting synchrotron radiation. In this case, starting from certain energy the 
synchrotron loss rate for the electrons becomes to dominate over ICS rate, which 
leads to suppression of the EM cascade development. Its penetration depth is 
then defined by the photon mean free path. Depending on the value of EGMF this 
transition may occur between $\sim 1\EeV$ and $\sim 10^6\EeV$.

 In this article we consider higher order process, double $e^+e^-$~pair 
 production (DPP) by photons.  
 The DPP cross section grows rapidly with $s$ near the threshold and quickly
 approaches the asymptotic value $\sigma (\infty) \simeq 6.45~\mb$ 
 \cite{Cheng:1970ef,Cheng:1970zb,Lipatov:1971ax}. 
 The explicit energy dependence 
 of the DPP cross section was estimated in Ref.~\cite{brown} by calculating 
 the dominant contribution from two $e^+e^-$ pairs to the absorptive part of 
 gamma-gamma forward scattering amplitude. 
 Since PP cross section decreases with the increase of $\sqrt{s}$ the DPP rate
 starts to dominate over PP rate above certain energy. For interactions with CMB the
 transition occurs above $\sim 1000\EeV$. In presence of the 
 radio background this energy goes up somewhat. If the EGMF is less than $10^{-11}\G$
 the EM cascade still exists at these energies and one should accurately count
 the secondary electrons from DPP. So far the EM cascade 
 simulations such as~\cite{propag, Lee:1996fp} roughly estimated DPP effect by 
 utilizing the total cross section and assuming that one $e^+e^-$~pair of the 
 two carries all the initial energy while two particles in the pair are produced
 with the same energy. By making use of CompHEP package~\cite{Boos:2004kh,Pukhov:1999gg,cite_comphep}
 we numerically calculate 
 differential cross section for DPP and compare the influence of DPP on 
 propagation of ultra high energy EM cascade with previous estimates.

The paper is organized as follows. In Sec.~\ref{CS} we present the results of the calculation 
of DPP cross section. In Sec.~\ref{TE} we write transport equations for EM cascade 
and calculate the coefficients for transport equations for photons, and secondary $e^+$,$e^-$ 
related to DPP. In Sec.~\ref{sec_models} we illustrate the influence of DPP in model example.
In Sec.~\ref{Conclusion} we summarize our results.

\section{DPP cross section calculation} \label{CS}
As it was mentioned in the Introduction the DPP process begins to dominate over 
PP at very high energies $E_{\gamma}\ga 1000$~EeV or $s \ga 1$~${\rm GeV}^2$, which is 
well beyond the DPP threshold. At these and higher energies DPP has 
noticeable effect on the propagation of EM cascade. In this energy region DPP 
total cross section is practically saturated by its asymptotic value. So, we 
are interested here mostly in energy and angular distributions of secondary 
electrons (positrons) in asymptotic regime ($s\to\infty$). 

We use CompHEP package for calculation of tree level differential DPP cross sections. 
This package allows to perform automatic calculations of matrix elements 
and their squares for any process $2\to 2$, .., $2\to 4$ at tree level. 
Then, with the aid of CompHEP one can integrate squared matrix elements 
over selected part of multi-particle phase space. See
Refs.~\cite{Boos:2004kh,Pukhov:1999gg,cite_comphep} for the details.

We introduce binning 
in the energy $E^*$ of one of the produced electrons\footnote{Here and further we denote 
by ``$^*$" quantities measured in the center of mass frame.}. Then we perform 
CompHEP simulations in the centre of mass frame (CMF) and obtain distributions 
over $\cos{\Theta^*}$ of the cross section in a given energy bin. Here 
$\Theta^*$ is the angle between the collision axis and the momentum vector of 
the electron.

\begin{figure}[htb]
\begin{center}
\includegraphics[width=0.7\columnwidth, height=0.5\columnwidth]{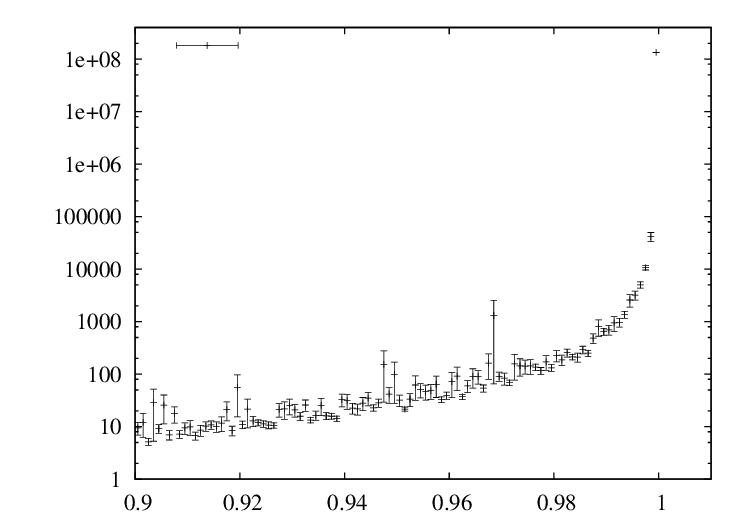} 
\put(-140.00,-5.00){\makebox(0,0)[cb]{{\small $\cos{\Theta^*}$}}} 
\put(-330.00,120.00){\makebox(0,0)[cb]{{\small $\sigma, pb$}}} 
\caption{\label{theta} Example of distribution of the DPP cross sections over 
$\cos{\Theta^*}$ for $\sqrt{s}=10.0$~GeV in energy $E^*$ bin $2.25-2.5$~GeV.}
\end{center}
\end{figure}

These calculations show that the angular distribution of secondaries tends to 
a strongly peaked function of $\cos{\Theta^*}$ in the asymptotic energy range. The 
peaks are located at forward and backward directions, i.e. at $\cos{\Theta^*}=\pm 
1$. This behavior is illustrated 
by Fig.~\ref{theta}
where we show an example of angular distribution for $\sqrt{s}=10.0$~GeV. 
The effect that most of secondaries go forward or backward becomes more pronounced 
with the increase of $\sqrt{s}$ and in the case of fixed 
$\sqrt{s}$ with higher energy $E^*$ of electron.
Numerically we found that the probability of emitting secondary electron inside
the cone with $ \mid 1 - \cos{\Theta^*} \mid < 1/50$ is 96.8\% for 
$\sqrt{s}=1.0$~GeV, 98.7\% for $\sqrt{s}=2.5$~GeV and 99.6\% for $\sqrt{s}=10.0$~GeV.
Also we checked that 
the probability of producing two forward secondaries of the same type (e.g., when both 
forward particles are electrons) integrated over energies and 
directions of other secondaries, is of order $10^{-3}$. So, the main part of events consists of two 
$e^+e^-$ pairs going to the opposite directions along the collision axis.

\begin{figure}[htb]
\begin{center}
\includegraphics[width=0.7\columnwidth, height=0.5\columnwidth]{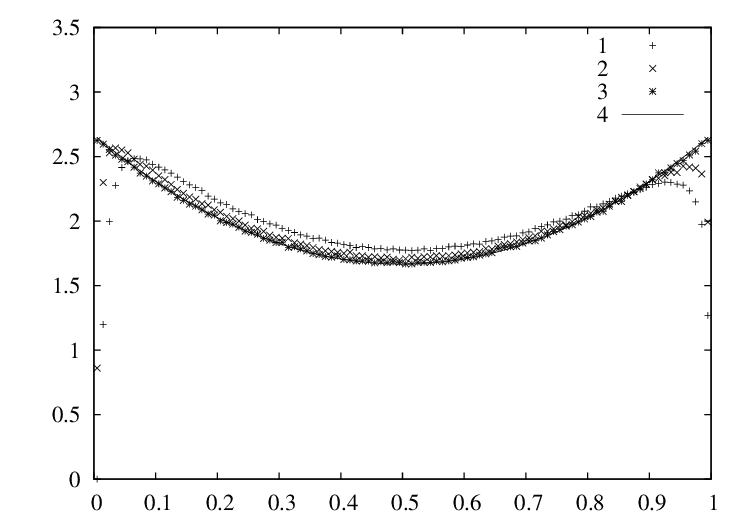} 
\put(-150.00,-5.00){\makebox(0,0)[cb]{{\small $r$}}} 
\put(-330.00,120.00){\makebox(0,0)[cb]{{\small $\phi(r,s)$}}} 
\caption{\label{energy_dis} The normalized energy distributions $\phi(r,s)$, 
$r=2E^*/\sqrt{s}$ (see main text for details): 1) $\sqrt{s}$ = 
$0.1$~GeV, 2) $\sqrt{s}$ = $0.25$~GeV, 3) $\sqrt{s}$ = $10.0$~GeV and 4) 
the analytic approximation Eq.~(\ref{fit}).}
\end{center}
\end{figure}

Let us now turn to the energy distribution. It is clear from the symmetry of the problem 
that the energy distribution in the CMF frame should be the same for the 
forward and backward electrons. Let us write the DPP differential cross section 
in the form

\begin{equation}
\frac{d\sigma}{dE^*} \equiv \frac{1}{\sqrt{s}} \phi 
\left(\frac{E^*}{\sqrt{s}/2},s\right)\sigma_{tot}(s)\;.\label{sigma}
\end{equation}
Then the energy conservation condition gives
\begin{equation}
\sqrt{s}\sigma_{tot}(s) = 4\int_0^{\sqrt{s}/2} E^*
\frac{d\sigma}{dE^*}\,dE^* = \sigma_{tot}(s)\,\sqrt{s}
\int_0^1 r\phi (r,s)\,dr \nonumber
\end{equation}
or for any value of $s$
\begin{equation}
\int_0^1 r\phi (r,s)\,dr = 1\;. \label{phiIntR}
\end{equation}
Imposing probability conservation requirement
\begin{equation}
\sigma_{tot}(s) = \int_0^{\sqrt{s}/2} \frac{d\sigma}{dE^*}\,dE^* \nonumber
\end{equation}
gives another integral constraint on $\phi(r,s)$:
\begin{equation}
\int_0^1 \phi (r,s)\,dr = 2\;. \label{phiInt}
\end{equation}
Although conditions~(\ref{phiIntR}) and~(\ref{phiInt}) do not necessary imply
${d\phi}/{ds}=0$ the results of CompHEP simulations show that for large 
enough $s$, when cross section approaches its asymptotic value, the energy 
distribution of secondaries in units of maximal energy $\sqrt{s}/2$ varies only 
slightly with $\sqrt{s}$. In Fig.~\ref{energy_dis}
we plot the distribution $\phi(r,s)$ as a function of $r$ for different values 
of $\sqrt{s}$. One can see that with varying $\sqrt{s}$ the only changes in 
these distributions are concentrated at the borders of the plot.
The distribution limit for $s\to\infty$ can be fitted (see fig.~\ref{energy_dis}) 
by simple analytic expression, which satisfies constraints 
(\ref{phiIntR}) and (\ref{phiInt}),
\begin{equation}
\label{fit}
\phi_{fit}(r) = \frac{5}{3} + (2r-1)^2.
\end{equation}
For comparison, the earlier approximation~\cite{propag, Lee:1996fp} in terms of the distribution
$\phi(r,s)$ reads
\begin{equation}
\label{phi_lee} 
\phi(r,s)=2\cdot\delta(r-0.5).
\end{equation}
In our further calculations we use for the energy distribution 
the Eq.~(\ref{sigma}), where $\phi(r,s) = \phi_{fit}(r)$ is given by Eq.~(\ref{fit})
and assume that all the secondary particles are directed alongside the collision axis.
In Sec.~\ref{sec_models} we discuss how good the above approximation is.

\section{Transport Equations}
\label{TE}
Here we describe propagation of the UHE cosmic rays using the formalism of 
transport equations in one dimension. Besides DPP term on which we are going
to focus now, the full transport equations for the electrons and photons contain
the terms describing ICS, PP, synchrotron and $e^+e^-$~pair production by electrons
and positrons
as well as redshift terms. For simplicity, we show below the part of the equation
written for nonexpanding universe with the terms related to the DPP process only.
\begin{eqnarray}
\frac{d}{dt} N_e (E_e,t) = \int_{E_e}^{\infty} dE_{\gamma}N_{\gamma}
(E_{\gamma},t) \int_{\epsilon_{min}}^{\epsilon_{max}} d\epsilon\,n(\epsilon) 
\int_{-1}^1 d\mu \frac{1-\mu}{2} \frac{d\sigma_{\rm DPP}}{dE_e} 
(E_e,E_{\gamma},s) \,\label{eeq}\\
\frac{d}{dt} N_{\gamma} (E_{\gamma},t) = - N_{\gamma} (E_{\gamma},t) 
\int_{\epsilon_{min}}^{\epsilon_{max}} d\epsilon\,n(\epsilon) \int_{-1}^1 d\mu
\frac{1- \mu}{2} \sigma_{\rm DPP} (s) \label{gammaeq}
\end{eqnarray}
where $N_e (E_e,t)$ is the (differential) number density of electrons at energy 
$E_e$ at time $t$, $n(\epsilon)$ is the number density of background photons at 
energy $\epsilon$, $\mu$ is the cosine of the collision angle
($\mu=-1$ for a head-on collision) and 
$s=2E_{\gamma}\epsilon(1- \mu)$ is center of mass energy squared. The term in 
the r.h.s. part of Eq.~(\ref{eeq}) describes influx of electrons produced 
in DPP. Transport equation for positrons has the same form as~(\ref{eeq}). 
The r.h.s. term of~(\ref{gammaeq}) describes the loss of 
photons due to DPP. The factor $(1- \mu)/2$ is the flux factor. 
As we've seen in the Sec.~\ref{CS} the pairs produced in DPP are directed alongside 
the collision axis. This implies that one of the pairs carries practically all the
initial energy of the photon in the laboratory frame. Here we neglect the nonleading pair
produced in the interaction.

Replacing integration over $\mu$ by integration over $s$ gives
\begin{eqnarray}
\frac{d}{dt} N_e (E_e,t) & = & \int_{E_e}^{\infty} dE_{\gamma}\frac{N_{\gamma}(E_{\gamma},t)}{8E_{\gamma}^2}
\int_{s_{th}}^{s_{max}} ds\,s \frac{d\sigma_{\rm DPP}}{dE_e}(E_e;E_{\gamma},s) 
I_{\epsilon}(\frac{s}{4E_{\gamma}}) ,\label{eeqs}\\
\frac{d}{dt} N_{\gamma} (E_{\gamma},t) & = & - \frac{N_{\gamma}(E_{\gamma},t)}{8E_{\gamma}^2}
\int_{s_{th}}^{s_{max}} ds\,s\, \sigma_{\rm DPP} (s) 
I_{\epsilon}(\frac{s}{4E_{\gamma}}) \label{gammaeqs}
\end{eqnarray}
where
\begin{equation}
I_{\epsilon}(x) = \int_x^{\epsilon_{max}}\frac{n(\epsilon)}{\epsilon^2}d\epsilon\;.
\label{inteps}
\end{equation}
Here $s_{th} = 16m_e^2$ is threshold CMF energy squared for DPP and $s_{max} = 
4E_{\gamma}\epsilon_{max}$.

Now we are ready to use the results obtained in the Sec.~\ref{CS}.
Again here we calculate the transport equation coefficients in the 
limit of $s{\gg}s_{th}$. This implies that electrons and positrons are 
ultrarelativistic in the CMF frame. The CMF $\gamma$-factor in the laboratory frame is 
\begin{equation}
\gamma_{CMF} \equiv (1-\beta_{CMF}^2)^{-\frac{1}{2}} = 
\frac{E_\gamma}{\sqrt{s}}\;. \nonumber
\end{equation}
Provided that $e^+$ and $e^-$ momenta are directed either towards the CMF frame 
velocity or in the opposite direction, their energy in the laboratory frame
\begin{equation}
E_e = \gamma_{CMF} E_e^* (1\pm \beta_e^*) = \frac{E_\gamma}{\sqrt{s}} E_e^* 
(1\pm \beta_e^*), \nonumber
\end{equation}
where $\beta_e^* \rightarrow 1$ is electron velocity in CMF. 
For the leading $e^+e^-$ pair we have:
\begin{equation}
E_e = 2 \frac{E_\gamma}{\sqrt{s}} E_e^*\;. \nonumber
\end{equation}
Then using Eq.~(\ref{sigma}) we finally obtain
\begin{equation}
\frac{d}{dt} N_e (E_e,t) = \frac{1}{16} \int_{E_e}^{\infty}
dE_{\gamma}\frac{N_{\gamma}(E_{\gamma},t)}{E_{\gamma}^3} 
\phi(\frac{E_e}{E_\gamma}) \int_{s_{th}}^{s_{max}} ds\,s\, \sigma_{\rm DPP}(s) 
I_\epsilon(\frac{s}{4E_\gamma})\;.\label{eeqs2}
\end{equation}
Using numerical simulations of cosmic rays propagation presented in the  
Sec.~\ref{sec_models} we have also verified that utilizing simple step function for the total 
cross section
\begin{equation}
\sigma_{\rm DPP}(s) = \sigma_{\rm DPP}(\infty) \Theta (s-s_{th})
\end{equation}
instead of exact one listed in~\cite{brown}, doesn't introduce any visible 
change to the resulting spectra. This implies that the equations~(\ref{eeqs2}) 
and~(\ref{gammaeqs}) can be simplified as follows:
\begin{eqnarray}
\frac{d}{dt} N_e (E_e,t) \simeq \sigma_{\rm DPP}(\infty) \int_{E_e}^{\infty}
dE_{\gamma}\frac{N_{\gamma}(E_{\gamma},t)}{E_{\gamma}} 
\phi(\frac{E_e}{E_\gamma}) K_\epsilon(s_{th}/4E_\gamma) ,\label{eeqs3}\\
\frac{d}{dt} N_\gamma (E_\gamma,t) \simeq - 2 \sigma_{\rm DPP}(\infty) 
N_{\gamma}(E_{\gamma},t) K_\epsilon(s_{th}/4E_\gamma) ,\label{gammaeqs3}
\end{eqnarray}
where
\begin{equation}
K_\epsilon(x) = \int_x^{\epsilon_{max}} I_\epsilon(y)y\,dy
\end{equation}
is the function totally determined by the background photons spectrum.

\section{Model example}
\label{sec_models}

In the previous sections we have found the precise expression for the 
distribution of secondary electrons from DPP. Here we consider a model 
example to illustrate the difference introduced by the specified cross section 
compared to the previous estimates.

We use a numerical code developed in Ref.~\cite{propag} to compute the flux of 
produced photons and protons. The code is based on the transport 
equations and calculates the propagation of nucleons, electrons and photons 
using the dominant processes. For EM cascade it includes all the processes 
mentioned above. For nucleons, it takes into account single and multiple pion 
production and $e^+e^-$ pair production, neutron $\beta$-decay. The 
propagation of  nucleons and the EM cascades are calculated self-consistently, 
that is secondary particles produced in all reactions are propagated alongside 
the primaries.
 
Besides CMB the radio, infra-red and optical (IRO) components of the universal 
photon background are taken into account in the simulation. Note that the radio 
background is not yet well known. Our results will depend strongly on the 
radio background assumed. Three models considered in this work are 
estimates by Clark {\it et al.}~\cite{clark} and the two models of Protheroe 
and Biermann~\cite{PB}, both predicting larger background than the first one. 
For the IRO background component we used the model~\cite{Stecker:2005qs}. This 
component doesn't have substantial effect on the propagation of UHE protons and 
EM cascade. For the strength of the random extragalactic magnetic field we use 
the range of values $10^{-12}{\rm G} < B < 10^{-11}{\rm G}$ following the 
estimate~\cite{dolag}.

Among the models we have chosen the one in which the UHE photons contribute substantial 
part of the total spectrum. Note that such models are strictly limited by the 
present experimental bounds on the photon component, see~\cite{Gelmini:2005wu} for details.
\begin{figure}[h]
\begin{center}
\includegraphics[height=0.75\textwidth,clip=true,angle=270]{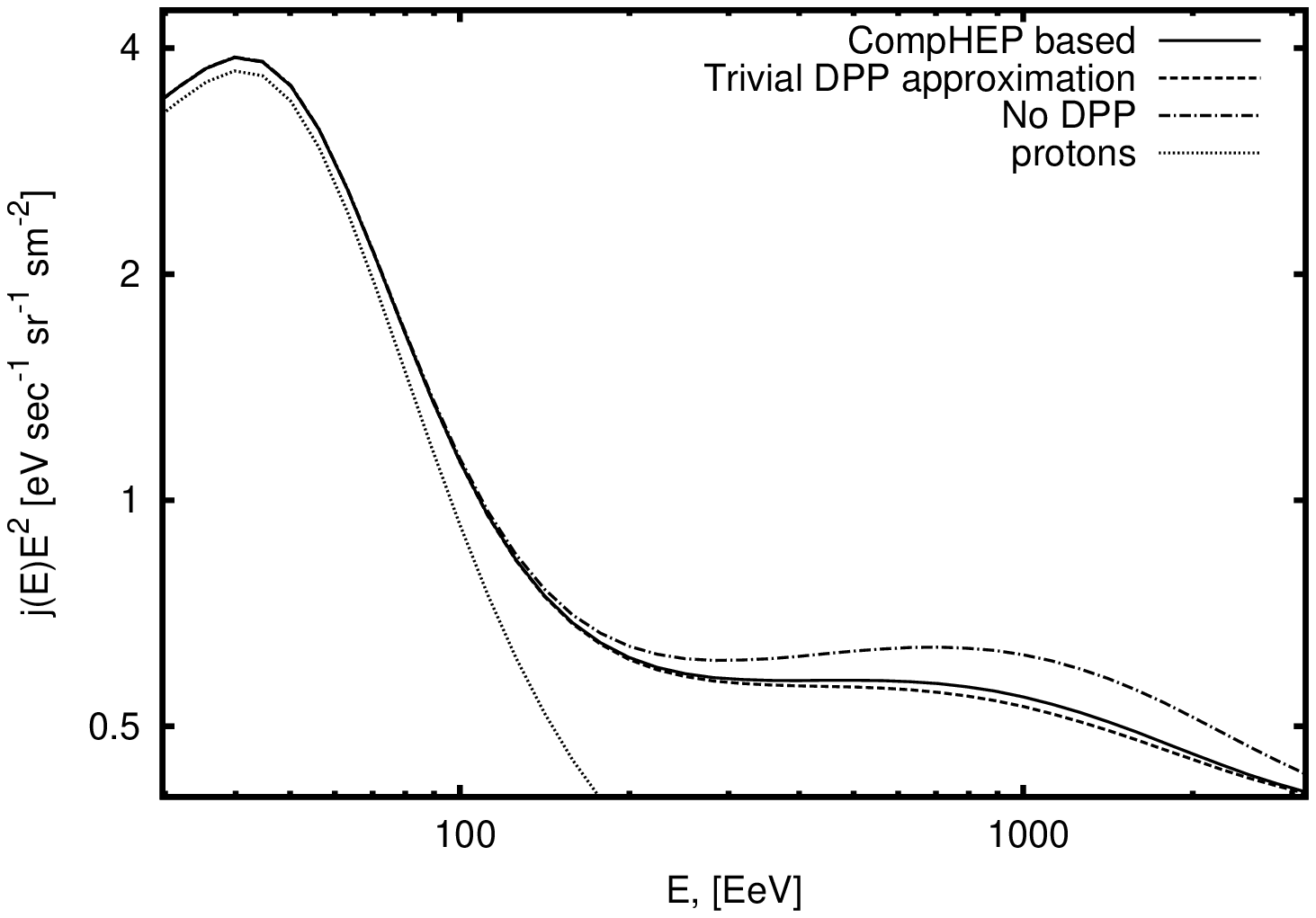} 
\caption[...]{Fluxes predicted by proton emitting source described in text. The 
dotted line represents proton component, while solid, dashed and dash-dotted 
lines represent total flux calculated with CompHEP based DPP, trivial DPP 
estimate and without DPP correspondingly.}
\label{gzk22}
\end{center}
\end{figure}

In Fig.~\ref{gzk22} the propagated cosmic ray flux is shown for proton sources with spectrum 
\begin{equation}
\frac{d\Phi}{dE} \sim E^{-1.5},\,\, E < 10^4 {\rm EeV}\
 \label{injection}
\end{equation}
homogeneously distributed in the Universe and having no evolution in the comoving frame.
The spectrum presented is normalized on HiRes~\cite{HiRes_spec} results
(fitting was done above $40\EeV$). The solid line represents the
total UHE cosmic ray flux calculated with use of new DPP estimate. The dotted 
line shows proton component. The dashed line shows total flux calculated using 
earlier DPP estimate (Eq.~\ref{phi_lee}), utilizing the total cross section and assuming that one 
$e^+e^-$~pair of the two carries all the initial energy while two particles in 
the pair are produced with the same energy. The dash-dotted lines are built without 
taking DPP into account at all. 

It is clear from the Fig.~\ref{gzk22} that DPP suppresses $\gamma$~ray flux above 
$100\EeV$. This is only true if the minimal radio background model~\cite{clark} 
is used. The same picture made for any of the two models of~\cite{PB} haven't 
shown any effect of DPP, since in this case the $\gamma$~flux is strongly 
suppressed by PP on radio. Increasing magnetic field above 
$10^{-11}{\rm G}$ also destroys the picture, this time due to synchrotron radiation. 
In the case of minimal radio background and moderate EGMF the trivial DPP effect 
estimate leads to extra suppression compared to the more accurate one proposed 
in this paper. Although overall error in terms of integral photon flux above 
$100\EeV$ turns to be only $+7$\% for the curve disregarding DPP and just $-1.5$\% 
for the trivial DPP estimate. Note that integral photon flux fraction predicted
in this model is 34\%, which is very close to the upper bound~\cite{A+Y}.
\begin{figure}[h]
\begin{center}
\includegraphics[height=0.75\textwidth,clip=true,angle=90]{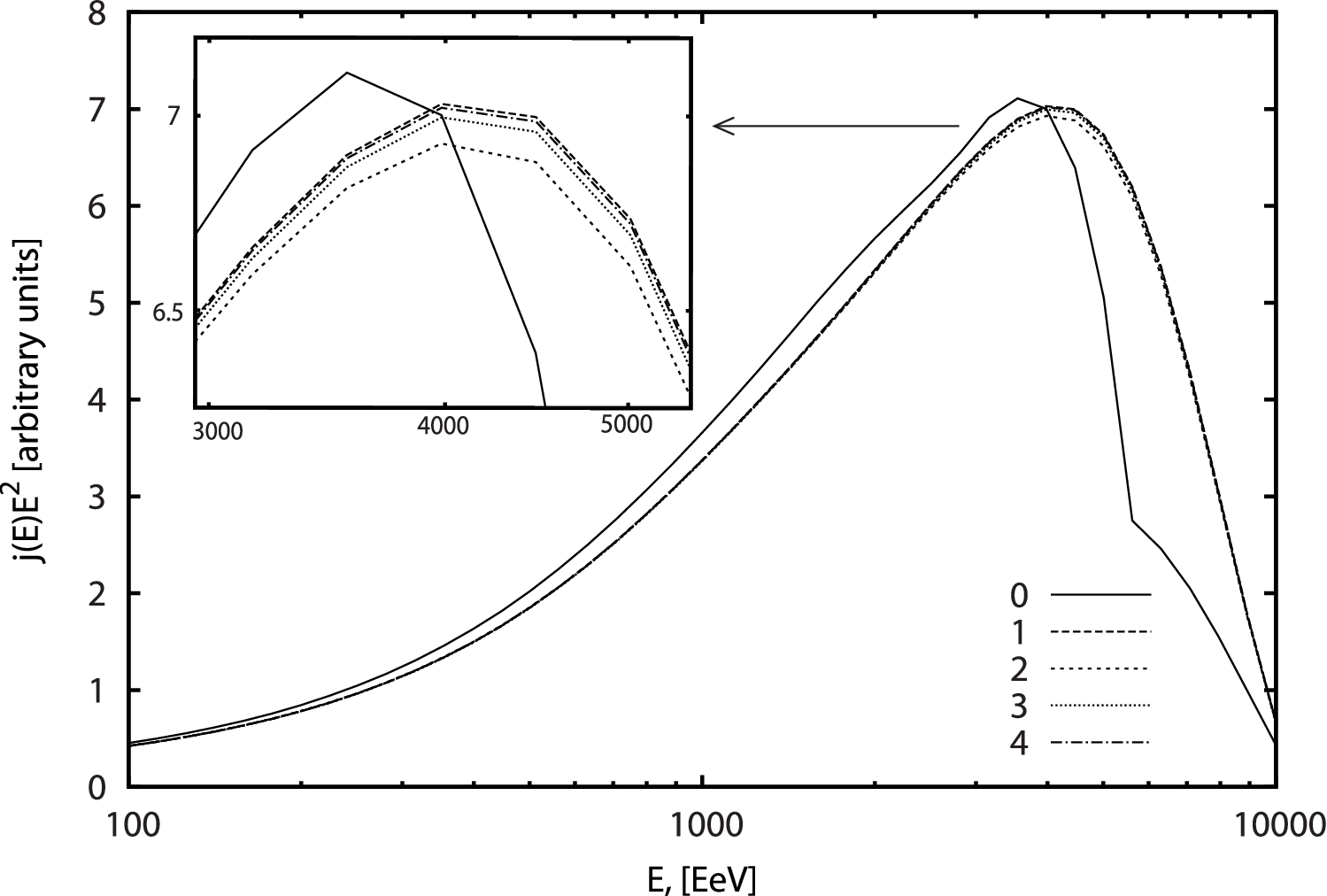} 
\caption[...]{Electron flux predicted by photon emitting source described in text using
0) earlier estimate Eq.~(\ref{phi_lee});
1) analytical fit~(\ref{fit}); 
2) analytical fit~(\ref{fit}) + 3.2\% perpendicular component (see details in text);
3) $\phi(r, 1\GeV^2)$;
4) $\phi(r, 100\GeV^2)$. }
\label{e500}
\end{center}
\end{figure}
So far we used the fixed energy distribution~(\ref{fit}) for all values of $s$. Also
we assumed that all the secondary particles are directed alongside the collision axis.
Let us now check how accurate the above approximations are. We have repeated our simulations
replacing the energy distribution~(\ref{fit}) by the tabulated functions obtained with
use of CompHEP for $\sqrt{s}=1\GeV$ and $\sqrt{s}=10\GeV$.
To see the maximal possible effect of the nontrivial angular distribution of secondaries
we have also repeated our calculations
assuming that 3.2\% of secondary particles are aligned perpendicular to the collision
axis in the CMF, while the rest of the particles are directed alongside the axis.
We don't show here the modified fluxes obtained in the model corresponding to Fig.~\ref{gzk22},
since they are practically indistinguishable from the curves already shown.
Instead to illustrate the maximal possible error introduced by the approximation used,
here we consider the pure photon sources with the same injection spectrum~(\ref{injection})
as in Fig.~\ref{gzk22} and count the income to the propagated electron and photon spectra from the uniformly
distributed sources located within $500\Mpc$ from the observer.
In Figs.~\ref{e500} and~\ref{ph500} the electron and photon fluxes in this model
are shown respectively.
\begin{figure}[h]
\begin{center}
\includegraphics[height=0.75\textwidth,clip=true,angle=90]{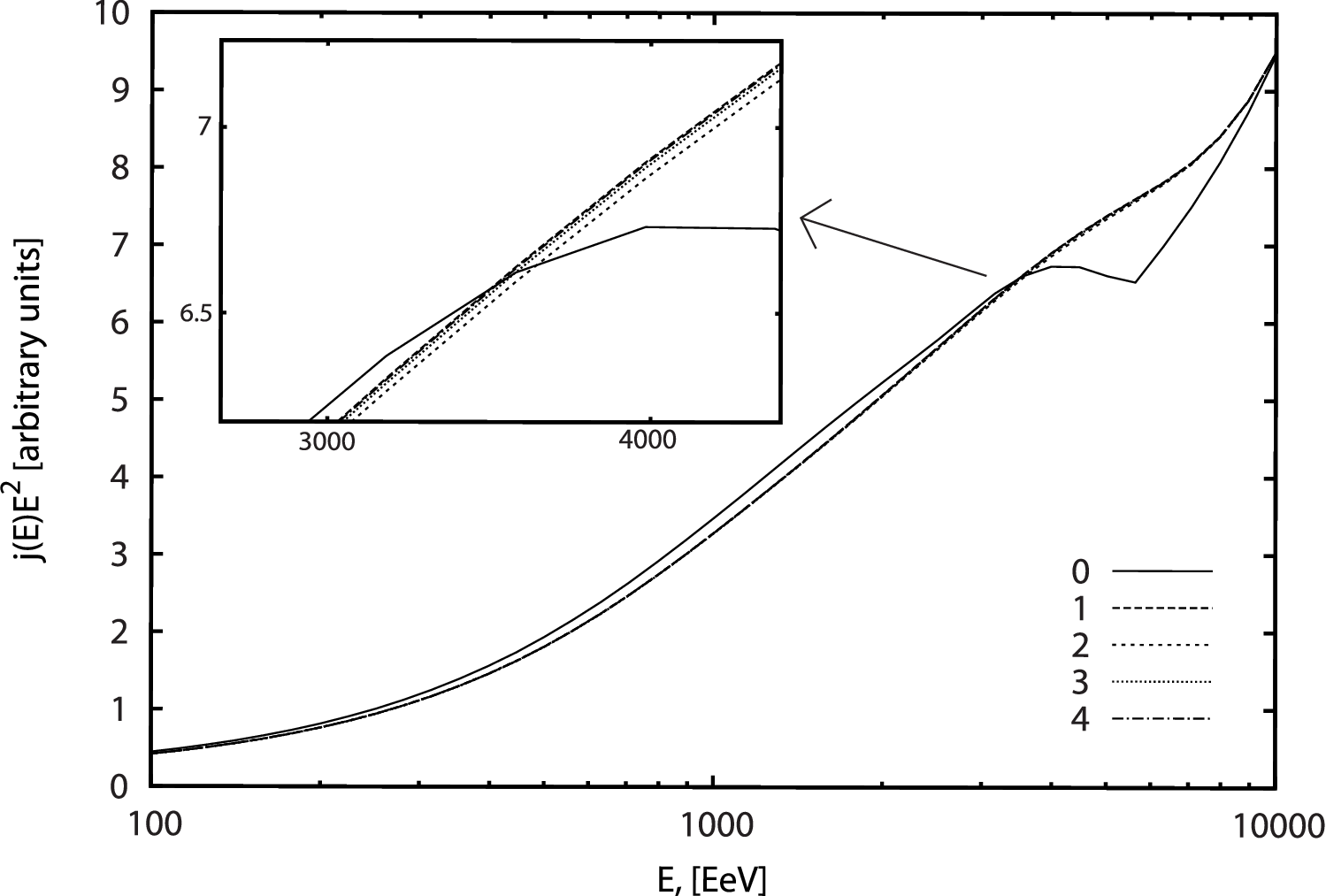} 
\caption[...]{Propagated photons flux predicted by photon emitting source described in text.
The designation of curves is the same as on Fig.~\ref{e500}. }
\label{ph500}
\end{center}
\end{figure}
Also on these figures the fluxes calculated using earlier estimate Eq.~(\ref{phi_lee}) are shown.
From the figures it is clear that the earlier estimate may lead to the artificial features in the spectra which
doesn't appear in our analysis. Also it is clear that discrepancy between the curves 1-4 representing different   
variants of our analysis are small compared to the error introduced by the earlier estimate. In fact the difference
between the curves 1-4 is comparable to the error introduced by finite energy binning used in our numerical code.

\section{Conclusion}
\label{Conclusion}

In this work we have considered in detail the DPP process. We have estimated 
the distribution of secondary electrons and positrons and made the improved 
cosmic rays simulation based on the new estimate. We have shown that in certain 
cases the DPP process may modify the photon component of the spectrum 
substantially. However this modification can only be seen if radio background 
is close to the minimal model~\cite{clark} and EGMF is lower than $10^{-11}{\rm G}$. 
In this case there is an energy range where DPP is the main attenuation 
mechanism for $\gamma$~rays and therefore differences in DPP estimates can 
clearly be seen. Although in the vast majority of the models which do not 
contradict to the present experimental bounds on the photon fraction in UHE 
cosmic rays DPP process doesn't make a substantial contribution to the 
attenuation and therefore can be treated simplistically or even disregarded.

We are indebted to D.~Gorbunov for inspiring us to prepare this 
work. We also thank D.~Semikoz for valuable discussions. This work was
supported in part by
the Russian Foundation of Basic Research grant 07-02-00820, by the grant of the
President of the Russian Federation NS-1616.2008.2, MK-1966.2008.2 (O.K.). Work 
of S.D. was supported in part by the grant of the Russian Science Support 
Foundation, Russian Foundation of Basic Research grants 08-02-00473-a and 
08-02-00768-a. Numerical part of the work was performed on the Computational 
cluster of the Theoretical Division of INR RAS.


\end{document}